%% file: main.tex
\documentclass[twocolumn,tighten]{aastex631}
\usepackage[starfontserif]{starfont}
\usepackage{graphicx,amsmath, natbib}
\usepackage{mathtools}
\usepackage[rightcaption]{sidecap}
\usepackage{appendix}
\usepackage{url, wrapfig, booktabs, enumitem}
\usepackage{orcidlink}

\DeclareSymbolFont{starfontsym}{OT1}{sts}{m}{n}
\DeclareMathSymbol{\mathSun}{\mathord}{starfontsym}{115}

\newcommand{\peff}{p_{\rm eff}}

\begin{document}
\title{An analytical galactic chemical evolution model with gas inflow and a terminal wind}

\author[0000-0002-7438-1059]{Kateryna A.\ Kvasova}
\affiliation{Department of Physics and Astronomy, University of Notre Dame, 225 Nieuwland Science Hall, Notre Dame, IN 46556, USA}
\author[0000-0001-6196-5162]{Evan N.\ Kirby}
\affiliation{Department of Physics and Astronomy, University of Notre Dame, 225 Nieuwland Science Hall, Notre Dame, IN 46556, USA}

\begin{abstract}
We present a new analytical galactic chemical evolution (GCE) model with gas inflow, internally caused outflow, and extra gas loss after a period of time. The latter mimics the ram pressure stripping of a dwarf satellite galaxy near the pericenter of its orbit around a host galaxy. The new model is called Inflow with Ram Pressure Stripping (IRPS). We fit the $\alpha$-element ([$\alpha$/H]) distributions of the Draco, Sculptor, Fornax, Leo~II, Leo~I, and And~XVIII dwarf spheroidal galaxies. We compared the best fits of IRPS with four other GCE models. The IRPS fits half of the galaxies in our set better than the Leaky Box, Pre-enriched, Accretion, and Ram Pressure Stripping models. Unlike previous models, none of the IRPS model parameters---not even the effective yield---correlates with galaxy properties, like luminosity.  One of the IRPS parameters is the $\alpha$-abundance at which stripping began.  That parameter can override the effective yield in determining the galaxy's mean $\alpha$-abundance.
\end{abstract}

\section{Introduction}
\label{sec:intro}

The past evolution of galaxies with simple star formation histories (SFHs) can be inferred with analytic one-zone GCE models \citep[][]{tal71,tin80}. This approach assumes instantaneous mixing, constant yields, and the instantaneous recycling approximation (IRA). All of these assumptions are violated in actual galaxies, in some cases severely.  Nonetheless, the models are practical, because they parameterize gas flows into and out of galaxies in ways that can make it easy to understand the processes that shape a galaxy's metallicity distribution function (MDF; in this work, we refer to all elements heavier than H and He as metals). Furthermore, their analytic solutions can be evaluated quickly, which means that the gas-flow parameters can be fit to observations efficiently.

Why do we need models with the IRA? Conceptually, they fit well for magnesium and other $\alpha$-elements, which are produced promptly by core-collapse supernovae. A limitation of the IRA lies in its inability to capture the chemical evolution of elements returned to the interstellar medium on longer timescales  \citep[e.g., iron;][]{brad03, mat08}. The separate delay times of $\alpha$ and Fe allow the {\rm [$\alpha$/Fe]} ratio to be used to infer the time span of chemical evolution. However, there is still important information that can be obtained through a study of a one-dimensional abundance distribution.  In particular, invoking the IRA is still appropriate for the $\alpha$-element distribution function (ADF)\@. Quantitative analysis of the ADF in the context of galactic gas inflow and catastrophic gas loss is the purpose of the present work. This framework can be applied especially to the many existing \citep[e.g.,][]{kir10, kir20, Ji20, woj21} and planned \citep[e.g., the Subaru Prime Focus Spectrograph,][]{tam18} large spectroscopic surveys that provide precise measurements of ADFs for the nearest dwarf spheroidal galaxies (dSphs). We will explore GCE models more appropriate for iron in a future work.

A well-known example of a GCE model is the Best Accretion Model of \citet{lyn75}, which can reproduce the MDF of the solar neighborhood, overcoming the G dwarf problem \citep{sch63}. This model presumes a functional form for gas inflow that leads to an analytic solution to the MDF. This inflow model might work well for the solar neighborhood, because the Milky Way is massive enough to keep attracting material. As a result, the gas expelled by stars is also likely to be returned to the system.

On the other hand, both gas inflow and gas loss are important in shaping the MDFs of dwarf galaxies \citep[e.g.,][]{kir11a,kir13, spe14}. To study dwarfs in the vicinity of a large host, a model with gas inflow, gas loss due to internal stellar feedback, and a terminal wind that approximates environmental effects should be developed.

In this Letter, we present a new one-zone model that introduces constant gas inflow into the two-phase GCE model, which is the Ram Pressure Stripping (RPS) model of \citet{kir13}. The model is called ``Inflow with Ram Pressure Stripping'' (IRPS). It assumes:
\begin{enumerate}[itemsep=0.5pt]
\item Instantaneous recycling and instantaneous mixing.
\item Monotonic metal abundance increase with time.
\item Kennicutt--Schmidt star formation law.
\item Internally caused gas outflow (feedback).
\item Constant pristine gas inflow before stripping (${\rm [\alpha/H]}<{\rm [\alpha/H]}_{\rm s}$). 
\item Sudden decrease of inflow to a minimal constant value and constant gas outflow (in addition to feedback-driven outflow) for ${\rm [\alpha/H]}>{\rm [\alpha/H]}_{\rm s}$\@.
\end{enumerate}
We draw heavily from \citet{pag09} for terminology (Table~\ref{tab:misc}). 



\input{misc.tab}

\section{Observational  Data} 
\label{sec:obs}
To test the IRPS model, we used the published abundances of member stars for Draco, Sculptor, Fornax, Leo~II, Leo~I \citep{kir13}, and And~XVIII \citep{kva24}. To obtain $\alpha$-element abundances [$\alpha$/H], we added the metallicity {\rm [Fe/H]} and $\alpha$-enhancement {\rm [$\alpha$/Fe]}, which were measured from Mg, Si, Ca, and Ti lines using the synthetic spectra method~\citep{kir08}.

Gaia proper motions were not available when \citeauthor{kir13}\ published their catalog of stars.  Therefore, we retroactively perform a membership cut based on the Gaia DR3 catalog \citep{gai16, gai23, gai23b}.  Most of these proper motions have high uncertainties (due to large distances), yet we used them to enhance the membership criteria. Stars with $>3\sigma$ deviation of {\rm $\mu_{\alpha *} = \mu_{\alpha} \cdot \cos{\delta}$}, {\rm $\mu_{\delta}$} within datasets (galaxy and nonmembers) were excluded. This has a negligible effect on the GCE model fitting, as most of the nonmembers were already excluded by other criteria \citep{kir13}.

\section{Methods} 
\label{sec:methods}

The GCE equations start with mass conservation:
\begin{equation}\begin{split}
\frac{dg}{dt}=F-E-\frac{ds}{dt}\Rightarrow\frac{dg}{ds}=\frac{F-E-ds/dt}{ds/dt}
\label{eq:first_main}
\end{split}\end{equation}
The model consists exclusively of gas and stars.  Other components, like dark matter, are presumed not to contribute to GCE\@. 

We invoke the Kennicutt--Schmidt law \citep{sch63}:
\begin{equation}
\frac{ds}{dt} = (1 - R)\psi = \lambda \psi = \beta g
\label{eq:sch}
\end{equation}

\noindent where $R$ is the return fraction from stellar deaths, $\lambda = (1-R)$, and $\beta$ is the star formation efficiency (SFE).

The galactic inflow {\rm $F$} is constant, and the gas-loss rate {\rm $E$} consists of a term for internal feedback (e.g., supernova-driven outflow), assumed to be proportional to the SFR, and a surplus gas leakage. \citet{gre03} discussed RPS as one of the most effective mechanisms for gas removal from dSphs, so we refer to the second term as RPS ({\rm $E'_{\rm s}$}):
\begin{equation}
E = \eta\beta g + E'_{\rm s}
\end{equation}
Thus, {\rm $E'_{\rm s}\neq0$} only after the moment when RPS commenced, ${\rm [\alpha/H]}>{\rm [\alpha/H]}_{\rm s}$. 

Finally, the total fraction of metals of gas $Z$ is introduced as \citep{pag09}
\begin{equation}
\frac{d(gZ)}{dS}=q+RZ-Z-Z_{\rm E}\frac{E}{\psi}+Z_{\rm F}\frac{F}{\psi},
\label{eq:second_main}
\end{equation}
In this work, $Z$ refers to the fraction of $\alpha$-elements only.

We recomputed the Leaky Box, Pre-enriched, Accretion, and RPS models' best-fitting parameters. Except for Accretion, the models are specific cases of IRPS: setting the inflow to zero yields the RPS model; setting the RPS parameter to zero yields the Pre-enriched model; and additionally setting the pre-enrichment to zero yields the Leaky Box model.

\subsection{Leaky Box Model}
\label{sec:lb}
The Leaky Box (Pristine) model tracks changes of the mass of the gas, the stars in the galaxy, and the gas $\alpha$-abundance \citep{sch63,tal71,sea72}. It assumes a primordial composition of the initial gas cloud and an outflow caused by internal processes, like stellar winds (``leaking''). Its ADF is
\begin{equation}
\frac{dN}{d{\rm [\alpha/H]}}\propto\left(\frac{10^{\rm [\alpha/H]}}{\peff}\right)\exp\left({\frac{-10^{\rm [\alpha/H]}}{\peff}}\right),
\label{eq:lb}
\end{equation}

\noindent where $p_{\rm eff}=p/(1+\eta)$ is the effective yield and $p$ is the true yield.

\subsection{Pre-enriched Model}
\label{sec:enri}
The Pre-enriched model assumes an enriched composition (${\rm [\alpha/H]}_0$) of the initial gas~\citep{pag09}:
\begin{equation}\begin{split}
\frac{dN}{d{\rm [\alpha/H]}}\propto\left(\frac{10^{\rm [\alpha/H]}-10^{{\rm [\alpha/H]}_{0}}}\peff\right)
\exp\left({\frac{-10^{\rm [\alpha/H]}}{\peff}}\right)
\label{eq:enri}
\end{split}\end{equation}

\subsection{Accretion Model}
\label{sec:accr}
The Accretion model \citep[the Best Accretion Model or the Extra Gas model;][]{lyn75} assumes a parameterized gas inflow. With a final-to-initial mass ratio $M$\@, $s$ is determined from
\begin{equation}\begin{split}
{\rm [\alpha/H]}(s) = \log\Biggl[\peff\left(\frac{M}{1+s-s/M}\right)^2\\
\times\left(\ln\left(\frac{1}{1-s/M}\right)-\frac{s}{M}\left(1-\frac{1}{M}\right)\right)\Biggr]
\label{eq:acc-s}
\end{split}\end{equation}
The ADF is
\begin{equation}\begin{split}
\frac{dN}{d{\rm [\alpha/H]}}\propto\left(\frac{10^{\rm [\alpha/H]}}\peff\right)\Biggl[1+s\left(1-\frac{1}{M}\right)\Biggr]\\
\times{\Biggl[\left(1-\frac{s}{M}\right)^{-1}-2\left(1-\frac{1}{M}\right)\left(\frac{10^{\rm [\alpha/H]}}{\peff}\right)\Biggr]}^{-1}
\label{eq:acc}
\end{split}\end{equation}
with $g=\Bigl(1+s\left(1-\frac{1}{M}\right)\Bigr)\cdot\Bigl(1-\frac{s}{M}\Bigr)$. 

\subsection{RPS Model}
\label{sec:rps}
Using three parameters, the RPS model  \citep{kir13} describes the evolution of a dwarf galaxy that lost its gas from an interaction with a large companion. Specifically, the motion through the  circumgalactic medium exerts a pressure on the dwarf galaxy, expelling its gas \citep{gg72, may06, zav12}. 

Before the RPS has started (${\rm [\alpha/H]}<{\rm [\alpha/H]}_{\rm s}$), the ADF is the Leaky Box. Later, the ADF is
\begin{equation}\begin{split}
\frac{dN}{d{\rm [\alpha/H]}}\propto\left(\frac{10^{\rm [\alpha/H]}}{\peff}\right) \Biggl[\exp\left({\frac{-10^{\rm [\alpha/H]}}{\peff}}\right)\\
+ \zeta \left(\exp\left({\frac{10^{{\rm [\alpha/H]}_{\rm s}}-10^{\rm [\alpha/H]}}{\peff}}\right)-1\right)\Biggr],
\label{eq:rps}
\end{split}\end{equation}
where $\zeta=E'_{\rm s}/\left(\beta(1+\eta)\right)$ is the ratio of the ram pressure gas loss to internally caused gas ejection.

\subsection{IRPS Model}
\label{sec:A2}

From Equations~(\ref{eq:first_main}) and (\ref{eq:second_main}), the ADF of IRPS is
\begin{gather}
\frac{dN}{d{\rm [\alpha/H]}}\propto\frac{g \cdot 10^{\rm [\alpha/H]}}{p_{\rm eff}+\left(10^{{\rm [\alpha/H]}_{\rm F}}-10^{{\rm [\alpha/H]}}\right)\cdot \xi/g},
\label{eq:irps}
\end{gather}
similar to the Accretion model. The constant inflow rate $F$ is normalized by internally caused gas loss as $\xi=F/\left(\beta(1+\eta)\right)$\@.

The gas fraction $g$ is found from Equations~(\ref{eq:first_main}) and (\ref{eq:second_main}). However, it is easier to present the $\alpha$-abundance as a function of gas rather than the reverse. Before the onset of RPS (${\rm [\alpha/H]}<{\rm [\alpha/H]}_{\rm s}$), the gas fraction is obtained from
\begin{eqnarray}
z &=& \frac{Z}{p} \notag \\
&=& \frac{\frac{F}{\beta (1+\eta)}- g}{g \cdot (1+\eta)} \ln\Biggl[\frac{\frac{F}{\beta (1+\eta)}-g}{\frac{F}{\beta (1+\eta)}-g_0}\Biggr]  + 
\frac{g_0}{g} \frac{\frac{F}{\beta (1+\eta)}-g}{\frac{F}{\beta(1+\eta)}-g_0} z_0+\notag \\
& & \frac{\frac{F}{\beta (1+\eta)} \cdot (g-g_0) }{g\cdot(\frac{F}{\beta(1+\eta)}-g_0)} \cdot \left(\frac{1}{1+\eta} + z_{\rm F}\right),
\label{eq:before}
\end{eqnarray}
which is equivalent to
\begin{eqnarray}
z' &=& z(1 + \eta) \notag \\
&=& \left(\frac{\xi-g}{g}\right) \ln\left(\frac{\xi-g}{\xi-g_0}\right) + \frac{g_0}{g} \left(\frac{\xi-g}{\xi-g_0}\right) z'_0 +\notag \\
& & \frac{\xi \cdot (g-g_0) }{g\cdot(\xi-g_0)} \left(1 + z'_{\rm F}\right),
\label{eq:before2}
\end{eqnarray}
where $g_0$ is the initial gas fraction, which we define to be unity, so that the mass is in units of $g_0$.

${\rm [\alpha/H]}_{\rm s}$ determines the duration of the accretive phase compared to the forced gas-loss phase rather than the peak of the ADF, in contrast to the RPS model of \citet{kir13}. After RPS has started,
\begin{gather}
z = \frac{\beta}{F} \left(g - \frac{F-E'_{\rm s}}{\beta (1+\eta)}\right){}_2 F_1 \Biggl[1, 1, \frac{F}{F-E'_{\rm s}}+1, \frac{g \beta (1+\eta)}{F-E'_{\rm s}}\Biggr] - \notag \\
\Biggl[\frac{g_0\left(g - \frac{F-E'_{\rm s}}{\beta (1+\eta)}\right)}{g \left(g_0-\frac{F-E'_{\rm s}}{\beta (1+\eta)}\right)}\Biggr]^{\frac{F}{F-E'_{\rm s}}} \cdot \Biggl\{\frac{\beta}{F}\left(g_0 - \frac{F-E'_{\rm s}}{\beta(1+\eta)}\right) \cdot \notag \\
{}_2 F_1 \Biggl[1, 1, \frac{F}{F-E'_{\rm s}}+1, \frac{g_0 \beta (1+\eta)}{F-E'_{\rm s}}\Biggr] + \notag \\
z_{\rm F} - z_0 + \frac{F-E'_{\rm s}}{F(1+\eta)}\Biggr\} + \frac{F-E'_{\rm s}}{F(1+\eta)} +z_{\rm F},
\label{eq:after}
\end{gather}
or
\begin{gather}
z' = \frac{g - \left(\xi - \zeta\right)}{\xi} \cdot {}_2 F_1 \Biggl[1,1, \frac{\xi}{\xi-\zeta} +1, \frac{g}{\xi-\zeta}\Biggr] -\notag \\
\Biggl[ \frac{g_0 \left(g - (\xi- \zeta)\right)}{g \left(g_0 - (\xi- \zeta)\right)}\Biggr] ^{\frac{\xi}{\xi - \zeta} } \cdot \Biggl\{ \frac{g_0- (\xi- \zeta)}{\xi} \cdot \notag \\
{}_2 F_1 \Biggl[1, 1, \frac{\xi}{\xi-\zeta} + 1, \frac{g_0}{\xi-\zeta}\Biggr] + \notag \\
z'_{\rm F} - z'_0 + \frac{\xi-\zeta}{\xi}\Biggr\} + \frac{\xi-\zeta}{\xi} + z'_{\rm F},
\label{eq:after2}
\end{gather}
where {\rm ${}_2 F_1(a, b, c, d)$} is a hypergeometric function (implemented with \texttt{scipy.special.hyp2f1}\footnote{ \texttt{WolframAlpha} was also used: \url{https://www.wolframalpha.com/}}
). {\rm $g_0$} and {\rm $z'_0$} are the gas fraction and $\alpha$-abundance for the beginning of the RPS epoch (the last point of Equation~(\ref{eq:before2})).

For the ${\rm [\alpha/H]}>{\rm [\alpha/H]}_{\rm s}$ case, if {\rm $E'_{\rm s}=F$} 
\begin{gather}
z' = e^{-\xi\left(\frac{1}{g}-\frac{1}{g_0}\right)} \left(z'_0 - z'_{\rm F} \right)+ z'_{\rm F}+\notag \\
 e^{-\frac{\xi}{g}}\Biggl[ Ei\left(\frac{\xi}{g}\right) - Ei \left(\frac{\xi}{g_0}\right)\Biggr],
\end{gather}
where {\rm $Ei()$} is an exponential integral.

Equation~(\ref{eq:before2}) is determined for $g\neq\xi$. If $g$ becomes equal to $\xi$, then  $g(t)={\rm Const}$, and the Extreme Inflow model applies~\citep{lar86}.  That model assumes a gas inflow rate that exactly balances the gas loss. We avoided these cases here. Table~\ref{tab:misc} (obtained from Equations (\ref{eq:first_main}) and (\ref{eq:second_main})) summarizes whether $g$ and $z$ increase or decrease for different conditions, and we also assumed that the $\alpha$-abundance only grows. So, before RPS has started, we used only solutions where the gas was less than the normalized inflow $\xi$. For the strong gas-loss phase, an intense influx of pristine gas may cause the formation of metal-poor stars. However, we do not exclude the inflow entirely, as the interaction with the environment may supply external gas for the infalling part, and during the RPS the dwarf galaxy can also re-accrete the ejected gas~\citep[e.g.,][]{may06}. From that, for ${\rm [\alpha/H]}>{\rm [\alpha/H]}_{\rm s}$\@, we set the inflow rate to a minimal constant $\xi=0.01$\@.

Instead of solving Equations (\ref{eq:before2}) and (\ref{eq:after2}), we derived {\rm $z'$} at each value in an evenly spaced array of $g$ values. Then we interpolated $z'(g)$ for 5001 {\rm [$\alpha$/H]} points between {\rm $-5$} and {\rm $0$} using \texttt{np.interp}. {\rm $g([\alpha/H])$} was constructed from Equations~(\ref{eq:before2}) (for {\rm $[\alpha/H]<[\alpha/H]_{\rm s}$}) and~(\ref{eq:after2}) ({\rm $[\alpha/H]>[\alpha/H]_{\rm s}$}). 
As the fixed range of metallicities that we used for the calculation is wider than any ADF, we enforced the constraint that below $z'(g=1)$ the gas fraction is unity and above $z'(g=0)$ the gas fraction is zero.\footnote{A Python version of the IRPS model can be found online: \dataset[10.5281/zenodo.14690724]{https://github.com/kkvasova/irps_model}.} 

\input{results_updated.tab}

\begin{figure}
\centering
\includegraphics[width=.995\linewidth]{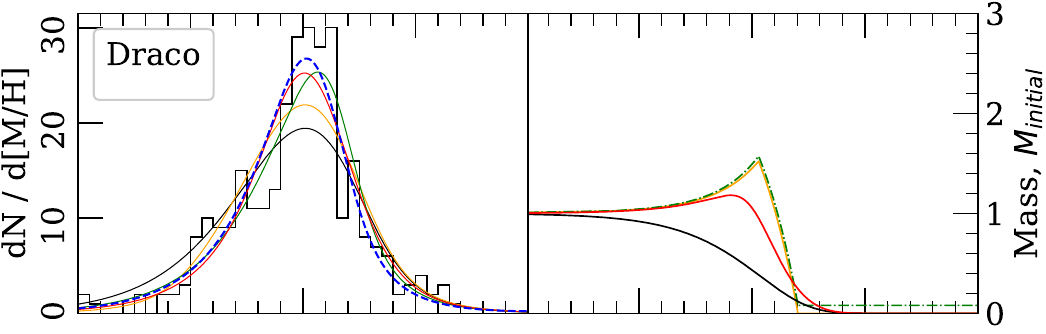}
\includegraphics[width=.995\linewidth]{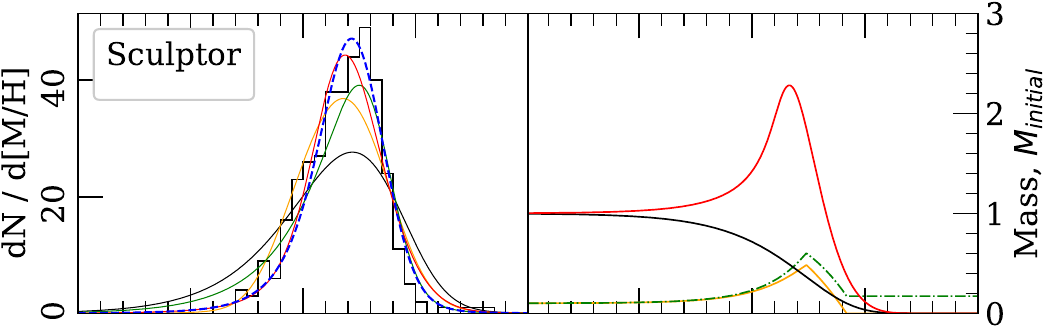}
\includegraphics[width=.995\linewidth]{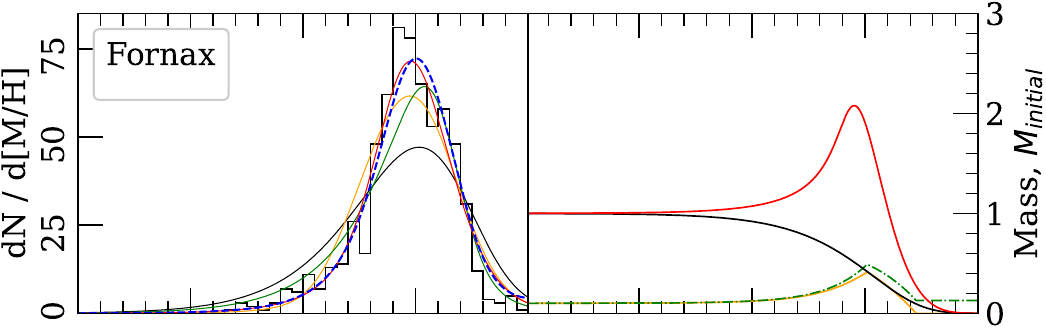}
\includegraphics[width=.995\linewidth]{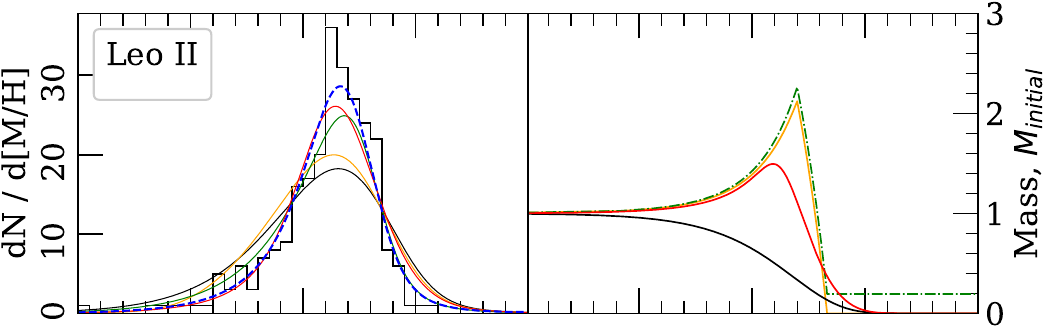}
\includegraphics[width=.995\linewidth]{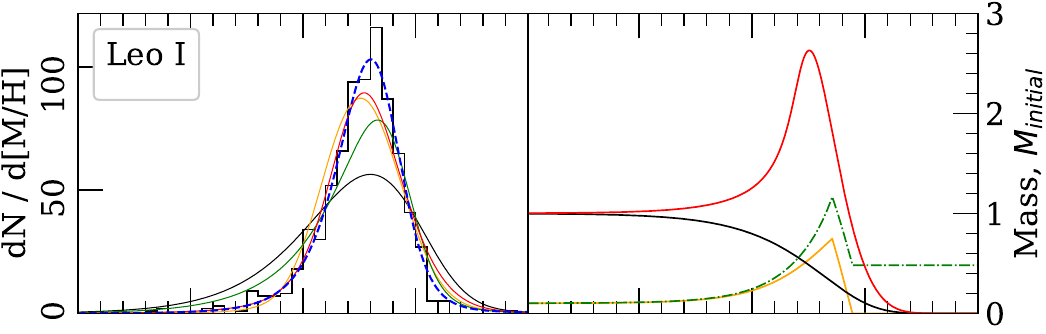}
\includegraphics[width=.995\linewidth]{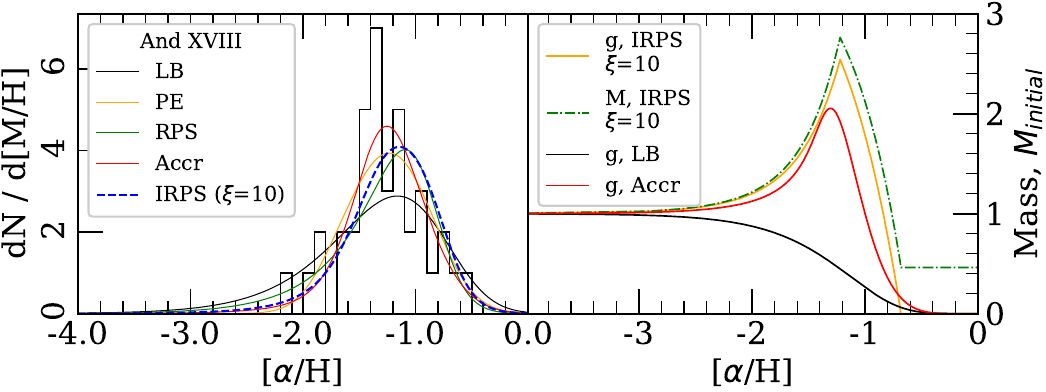}
\caption{
ADFs and the best-fitting GCE models for the six dwarf galaxies. There are Leaky Box (black), Pre-enriched (orange), RPS (green), Accretion (red), and IRPS (blue dashed) models over histograms of the member stars (black). The right panels illustrate the gas fraction (orange), the total mass from the IRPS model (green; we fix the mass-loading factor as $\eta=0.5$; the RPS begins at the inflection points in these curves), and the Accretion (red) and Leaky Box (black) gas laws. The units of mass are the initial gas mass of the galaxy, although IRPS mass functions that do not start at unity were divided by initial $\xi$. 
}
\label{fig:gcem}
\end{figure}

\subsection{Likelihood for Chemical Evolution Models}

To find the most likely parameters of each model, we maximized the logarithm of the likelihood, which was constructed as follows:
\begin{equation}\begin{split}
L=\prod_i \int_{-\infty}^{\infty} \frac{dP}{d{\rm [\alpha/H]}} \frac{1}{\sqrt{2\pi}\cdot\delta{\rm [\alpha/H]}_i}\\
\times\exp{\left(-\frac{\left({\rm [\alpha/H]}-{\rm [\alpha/H]}_i\right)^2}{2\left(\delta{{\rm [\alpha/H]}}_i\right)^2}\right)}\,d{\rm [\alpha/H]},
\label{eq:5}
\end{split}\end{equation}
where {\rm [$\alpha$/H]$_i$} is the measured abundance of each star $i$, and $\delta{{\rm [\alpha/H]}}_i$ is its uncertainty.

The logarithm of $L$ was maximized with an ensemble Markov Chain Monte Carlo (MCMC) sampling of the space of the parameters implemented by the \texttt{emcee} Python library (Section~\ref{sec:res}).

For an $\alpha$-abundance error, it is required to use the iron measurement error. This includes the systematic error and a random uncertainty:
\[\begin{split}
\delta {\rm [Fe/H]}_i=\sqrt{\delta {\rm [Fe/H]}_{i,{\rm rand}}^2+\delta{\rm [Fe/H]}_{\rm sys}^2},
\end{split}\]

\noindent where $\delta{\rm [Fe/H]}_{i,{\rm rand}}$ is from the spectral fit, and $\delta{\rm [Fe/H]}_{\rm sys} = 0.106$ \citep{kir10,kir15c}.

[$\alpha$/H] abundances were obtained as the sum of [Fe/H] and [$\alpha$/Fe], and we assumed the same systematic error as for [Fe/H]\@. So the total error is
\[\begin{split}
\delta {\rm [\alpha/H]}_i=\sqrt{\delta {\rm [\alpha/Fe]}_{i, {\rm rand}}^2+ \delta {\rm [Fe/H]}_{i,{\rm rand}}^2+\delta{\rm [Fe/H]}_{\rm sys}^2}
\end{split}\]

To compare the models, we used the corrected Akaike information criterion \citep[AICc;][]{aka74,sug78}:

\begin{equation}
{\rm AICc} = -2 \ln L + 2r + \frac{2r(r+1)}{N-r-1}
\end{equation}
where $L$ is the likelihood, $r$ is the number of model parameters (1, 2, 3, 2, and 3 for Leaky Box, Pre-enriched, RPS, Accretion, and IRPS, respectively), and $N$ is the number of stars. The smaller the AICc, the better the model.  The AICc penalizes additional free parameters, such that simpler models are sometimes preferred, even if their likelihood is smaller than a model with more free parameters.

\input{targets.tab}

\section{Results}
\label{sec:res}

The best-fitting GCE models are given in Table~\ref{tab:results} and Figure~\ref{fig:gcem}. After the priors were calculated with \texttt{numpy.random.uniform}, $10$ MCMC trials per 1000 (10 for Accretion) steps were iterated for all but the IRPS and RPS models.
We used 30, 5, 30, and 10 chains, each with $3\cdot10^4$, $10^4$, $10^5$, and $10^5$ iterations, for the MCMCs for the Leaky Box and Pre-enriched, Accretion, RPS, and IRPS models, respectively. We excluded 10$\%$ (30$\%$ for Accretion) of the iterations. All terms are consistent with \citet{kir11b, kir13}, except for the RPS model, possibly due to our refined membership criteria.

The parameter $\xi$ in the IRPS model can be considered as a normalization factor, so it can be assumed as some constant value. We found the best fits for three parameters of the IRPS model: $p_{\rm eff}$, $\zeta/\xi$ (treated as a single parameter with $\xi=10={\rm Const}$), and [$\alpha$/H]$_{\rm s}$\@. We determined $p_{\rm eff}$, which incorporates the wind mass-loading term $\eta$ and the yield $p$.  In principle, the mass-loading term could be determined if the yield were known or assumed (from an initial mass function and theoretical nucleosynthesis calculations). The initial conditions are {\rm $g_0 = 1$}, {\rm ${\rm [\alpha/H]}_{\rm F} = {\rm [\alpha/H]}_0 = -5$}, and {\rm $g_0 = g_{\rm s}$}, {\rm [$\alpha$/H]$_{\rm F}=-5$}, {\rm [$\alpha$/H]$_0 = $ [$\alpha$/H]$_{\rm s}$} for Equations~(\ref{eq:before2}) and (\ref{eq:after2}), respectively.


For the ADF, the previous best-fitting models were replaced with the IRPS model for Draco, Leo~II, and Leo~I\@. The sample size of And~XVIII is small, so the AICc prefers simpler models. Its ADF is similar to the [Fe/H] distribution function of Ursa Minor. Both of them have a metal-rich tail, which is unusual for dSphs.  It is possible that these galaxies lost their metal-poor stars in tidal disruption \citep[e.g.,][]{kir11b, kir11a}.

The Accretion model approximates the ADFs of most dSphs well. The IRPS model fits even better for most dSphs. Furthermore, the final masses from the IRPS model (see Figure~\ref{fig:gcem}, Section~\ref{sec:zt}) are higher (lower) than the initial gas masses of galaxies whose former best-fitting model was Accretion (RPS) for most dSphs (except And~XVIII)\@. This correlation demonstrates the large importance of environment in dictating the fate of a dwarf galaxy. External processes, like accretion or stripping, determine the galaxy's most fundamental parameter: mass.

\section{Discussion}
\label{sec:discuss}

\subsection{Model Parameter Dependence on Galaxy Properties}

\begin{figure}[h!]
\centering
\includegraphics[width=1.\linewidth]{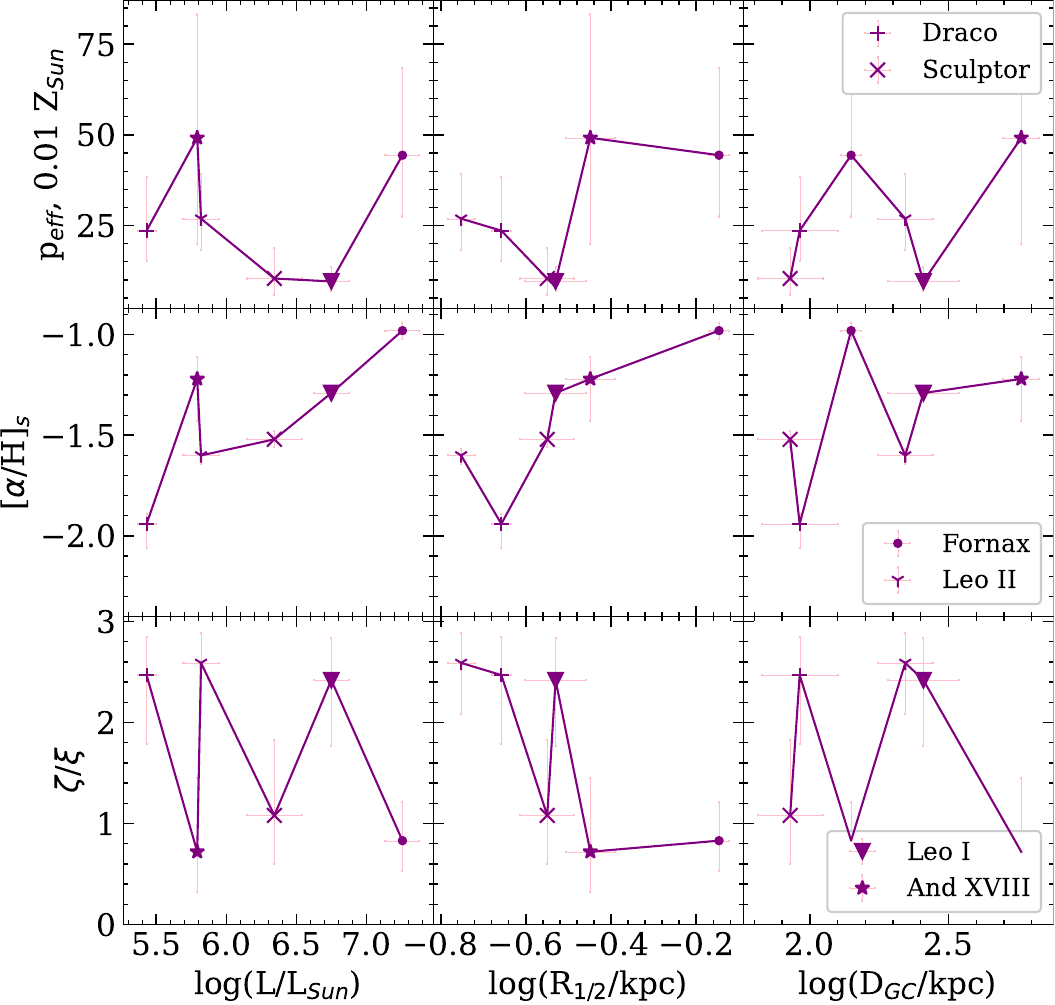}
\caption{Results of the IRPS model: effective yield ($p_{\rm eff}$), $\alpha$-abundance at which stripping commences ({\rm [$\alpha$/H]$_{\rm s}$}),  and outflow-to-inflow ratio ({\rm $\zeta/\xi$}) vs.\ the galaxies' luminosities, half-light radii, and distances from the host.}
\label{fig:trends}
\end{figure}

Figure~\ref{fig:trends} shows the dependence of IRPS parameters on luminosity, half-light radius, and distance from the host galaxy (Table~\ref{tab:targets}).  There are no clear trends.  While the chemical evolution model results (Section~\ref{sec:res}) reflect the strong influence of environment on chemistry, $D_{\rm GC}$ does not appear to be a good indicator of the strength of that environment. \citet{kir11a} also found that $D_{\rm GC}$ is a poor predictor of chemical evolution.  The current position of a satellite is just one snapshot in time, and it is biased toward the apocenter, which is a poorer metric of environmental influence than pericenter.

The effective yield was the only chemical evolution parameter that \citet{kir11a} found to correlate with any galaxy property, which was luminosity.  This correlation is essentially the mass--metallicity relation \citep{kir13}, where $p_{\rm eff}$ and $L$ are proxies for metal abundance and mass.  That correlation disappears in the IRPS model. 
${\rm [\alpha/H]}_s$ can control the mean $\alpha$-abundance of the galaxy as well as $p_{\rm eff}$. Whereas $p_{\rm eff}$ reflects the depth of the potential well, a low value of ${\rm [\alpha/H]}_s$ can lower the mean $\alpha$-abundance, overriding the influence of $p_{\rm eff}$.  Our result might help to explain how dSphs have a large range of mean metallicities, despite having similar depths of their gravitational potentials \citep{str08}.

\subsection{Age--$\alpha$-abundance Relations}
\label{sec:zt}

\begin{figure}[h!]
\centering
\includegraphics[width=1.\linewidth]{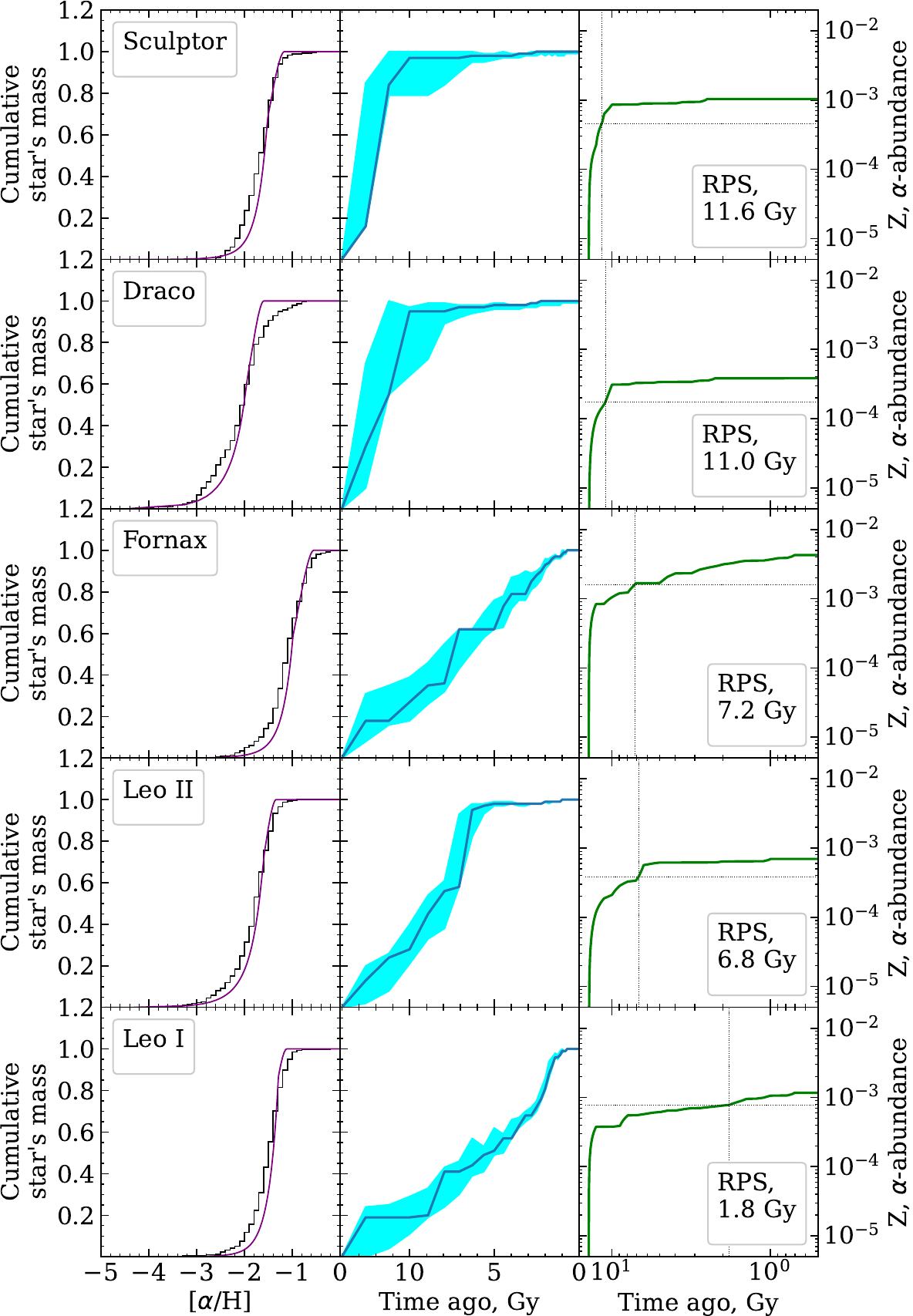}
\caption{Left two columns: cumulative stellar mass as a function of [$\alpha$/H] and lookback time.  Right column: age--$\alpha$-abundance relations from~\citet[][{\rm $Z(t) = Z_{\mathSun} 10^{\rm [\alpha/H](t)}$}]{wei14_2}. The dotted lines mark the beginning of RPS\@.}
\label{fig:z_evol}
\end{figure}

We deduced the age--$\alpha$-abundance relations $Z(t)$ for five galaxies by combining the $s(Z)$ from the {\rm [$\alpha$/H]} IRPS model with the observed SFHs \citep{wei14_1}.  Figure~\ref{fig:z_evol} illustrates the procedure to obtain $Z(t)$\@.

$s(Z)$ is the cumulative integral of the ADF, but it can also be found by the difference between the total mass $M(Z)$ and $g(Z)$ (Figure~\ref{fig:gcem}):
\begin{gather}
\frac{dM / dt}{dg / dt} = \frac{F-E}{F-E-ds/dt}\rightarrow\frac{dM}{dg} = \frac{F-E}{F-E - \beta g}
\end{gather}

For ${\rm [\alpha/H]}>{\rm [\alpha/H]}_{\rm s}$ the mass is
\begin{equation}\begin{split}
M =\frac{\eta \left( g - g_0\right)}{1+\eta}  -\frac{\xi - \zeta}{1+\eta} \ln \Biggl[ \frac{\xi - \zeta -g}{\xi -\zeta - g_0}\Biggr]+ M_{0}
\label{eq:mass_after}
\end{split}\end{equation}

\noindent Before RPS (${\rm [\alpha/H]}<{\rm [\alpha/H]}_{\rm s}$), the same equation is used, with {\rm $g_0=M_0=1$}, {\rm $\zeta = 0$}.  The final mass value before stripping determines the initial conditions for the RPS epoch.

From Figure~\ref{fig:z_evol}, RPS began at $t_s = 11.0$, 11.6, 7.2, 6.8, and 1.8 Gy ago for Draco, Sculptor, Fornax, Leo~II, and Leo~I\@. The times {\rm $t_{\rm s}$} are consistent with the time when $\sim50\%-80\%$ of stars were born. Although we have referred repeatedly to RPS, we cannot conclude whether there was some externally caused gas loss, which then limited the total number of stars born, or whether the additional gas loss $\zeta$ was the result of internal processes. \citep{dol02, sohn13, sohn17, wei14_1, wei14_2}. 

The age--$\alpha$-abundance relation shows how the SFH affects metal enrichment (Figure~\ref{fig:z_evol}). To illustrate this point, we consider the functions {\rm $g(Z), s(Z)$} from the IRPS best-fitting models. We use the observed SFH {\rm $s(t)$} to infer $Z(t)$.\footnote{The spectroscopic ADFs used here differ from the MDFs inferred from the color-magnitude diagram \citep{wei14_2}.  The spectroscopic ADFs should be more accurate.} Then, we interpret $Z(t)$ in the context of classic numerical GCE models obtained from the Kennicutt-Schmidt star formation law and the closed-box GCE\@. \citet[][their Figure~2]{tal71} concluded that $Z(t)$ is shallower when the SFE (called {\rm $\beta$} here but called $\nu$ by \citeauthor{tal71}) is smaller. This situation can be treated with the IRA, because continuous star formation retains the faster core-collapse supernovae as the main drivers of metal enrichment. On the contrary, fast initial metal enrichment occurs for high SFE\@. High SFE implies bursty star formation, so any late enrichment is dominated by delayed type Ia supernova explosions.

Our results show that larger and distant dwarfs (Fornax and Leo~I) have significant recent metal enrichment from a steadily increasing $Z(t)$\@. For Draco, Sculptor, and Leo~II, the most important increase of the metal content was achieved very early. We conclude that the IRA assumption of our analytic model is more valid for Fornax and Leo~I than the smaller, more rapidly forming dSphs.

\section{Summary} 
\label{sec:concl}

We have derived the one-zone IRPS model from the basic principles of GCE\@. The best-fitting model parameters were found for Draco, Sculptor, Fornax, Leo~II, Leo~I, and And~XVIII\@. We compared our results for the ram pressure stripped galaxies with those from the accretive galaxies. This new model contributes to the body of analytic GCE models, by allowing for more possibilities for simultaneous gas inflow and outflow. The qualitative consistency of $Z(t)$ with the~\citealt{tal71} numerical results allows the usage of the model with some caution not only for ADF and $g(Z)$, but also for a rough estimate of when the gas available for star formation decreased to some critical value.

\begin{acknowledgments}
{

We thank the anonymous referee and the editor for a careful review, which enhanced the paper. Some of the data presented herein were obtained at Keck Observatory, which is a private 501(c)3 non-profit organization operated as a scientific partnership among the California Institute of Technology, the University of California, and the National Aeronautics and Space Administration. The Observatory was made possible by the generous financial support of the W. M.\ Keck Foundation. The authors wish to recognize and acknowledge the very significant cultural role and reverence that the summit of Maunakea has always had within the Native Hawaiian community. We are most fortunate to have the opportunity to conduct observations from this mountain.

This work has made use of data from the European Space Agency (ESA) mission {\it Gaia} (\url{https://www.cosmos.esa.int/gaia}), processed by the {\it Gaia} Data Processing and Analysis Consortium (DPAC; \url{https://www.cosmos.esa.int/web/gaia/dpac/consortium}). Funding for DPAC has been provided by national institutions, in particular the institutions participating in the {\it Gaia} Multilateral Agreement.

K.A.K. is deeply indebted to the Armed Forces of Ukraine for the defense of her family, friends, and people in wartime.
}
\end{acknowledgments}

\facility{Keck: II (DEIMOS)}

\software{astropy \citep{ast13, ast18, ast22}, corner \citep{for16}, emcee \citep{for13}, matplotlib \citep{hun07}, numpy \citep{van11}, scipy \citep{vir20}}

\bibliography{main}{}
\bibliographystyle{aasjournal}

\end{document}

%% file: misc.tab
\begin{deluxetable*}{cccccccccc}
\label{tab:misc}
\tablecolumns{10}
\tablecaption{GCE Model Parameters.}
\tablehead{ \multicolumn{10}{c}{Terms Used for the GCE Models\tablenotemark{a}} \\
\hline
Symbol & \multicolumn{2}{c}{Description and Units} & \multicolumn{3}{c}{Symbol} & \multicolumn{4}{c}{Description and Units}}
\startdata
$\lambda$      & \multicolumn{2}{c}{Locked-in-stars mass fraction}             & \multicolumn{3}{c}{$R=1-\lambda$}   & \multicolumn{4}{c}{Return mass fraction} \\
$S$           & \multicolumn{2}{c}{Total mass of stars born ($M_{\rm i}$)}     & \multicolumn{3}{c}{$s=\lambda S$}   & \multicolumn{4}{c}{Mass of existing stars ($M_{\rm i}$)} \\
$p$           & \multicolumn{2}{c}{True yield $\left(\frac{1}{\lambda}\right)$}               & \multicolumn{3}{c}{$q = \lambda p$} & \multicolumn{4}{c}{Fraction of newly created elements} \\
$g$           & \multicolumn{2}{c}{Mass fraction of gas ($M_{\rm i}$)}           & \multicolumn{3}{c}{$p_{\rm eff}=\frac{p}{1+\eta}$} & \multicolumn{4}{c}{Effective yield $\left(\frac{1}{\lambda}\right)$} \\
$F$           & \multicolumn{2}{c}{Gas inflow rate $\left(\frac{M_{\rm i}}{\rm Gy}\right)$}          & \multicolumn{3}{c}{$\xi=\frac{F}{\beta(1+\eta)}$} & \multicolumn{4}{c}{Normalized gas inflow rate $\left(\frac{M_{\rm i}}{{\rm Gy} \cdot \beta}\right)$} \\
$E$           & \multicolumn{2}{c}{Total gas outflow rate $\left(\frac{M_{\rm i}}{\rm Gy}\right)$}   & \multicolumn{3}{c}{$\zeta=\frac{E'_{\rm s}}{\beta(1+\eta)}$} & \multicolumn{4}{c}{Normalized terminal wind rate $\left(\frac{M_{\rm i}}{{\rm Gy} \cdot \beta}\right)$} \\
$E'_{\rm s}$  & \multicolumn{2}{c}{Terminal wind rate $\left(\frac{M_{\rm i}}{\rm Gy}\right)$}  & \multicolumn{3}{c}{$Z_{\rm s}=Z_{\mathSun} 10^{{\rm [\alpha/H]}_{\rm s}}=z_{\rm s} p=z'_{\rm s} p_{\rm eff}$} & \multicolumn{4}{c}{$\alpha$-abundance when RPS occurs} \\
$\eta$        & \multicolumn{2}{c}{Mass-loading factor}                          & \multicolumn{3}{c}{$Z=Z_{\mathSun} 10^{\rm [\alpha/H]}=z p=z' p_{\rm eff}$} & \multicolumn{4}{c}{Gas-phase $\alpha$-abundance of the system} \\
$M$           & \multicolumn{2}{c}{Mass of the system ($M_{\rm i}$)} & \multicolumn{3}{c}{$Z_{0}=Z_{\mathSun} 10^{{\rm [\alpha/H]}_{0}}=z_{0} p=z'_{0} p_{\rm eff}$} & \multicolumn{4}{c}{Initial gas-phase $\alpha$-abundance} \\
$\psi$        & \multicolumn{2}{c}{Star formation rate $\left(\frac{M_{\rm i}}{\rm Gy}\right)$}  & \multicolumn{3}{c}{$Z_{\rm E}=Z_{\mathSun} 10^{\rm [\alpha/H]}=z p=z' p_{\rm eff}$} & \multicolumn{4}{c}{$\alpha$-abundance of the outflow ($Z_{\rm E}=Z$ here)} \\
$\beta$       & \multicolumn{2}{c}{SFE ($\frac{1}{\rm Gy}$)}   & \multicolumn{3}{c}{$Z_{\rm F}=Z_{\mathSun} 10^{{\rm [\alpha/H]}_{\rm F}}=z_{\rm F} p=z'_{\rm F} p_{\rm eff}$} & \multicolumn{4}{c}{$\alpha$-abundance of the inflowing gas} \\
\hline
\hline
\multicolumn{10}{c}{Conditions for the IRPS Model\tablenotemark{b}}\\
\hline
\multicolumn{3}{c}{{\rm $g$}} & \multicolumn{2}{c}{{\rm $z<z_{\rm cr}$}} & \multicolumn{3}{c}{{\rm $z>z_{\rm cr}$}} & \multicolumn{2}{c}{{\rm $z_{\rm cr}= z_{\rm F}+\frac{g}{(1+\eta)\xi}$}} \\
\hline
\multicolumn{3}{c}{{\rm $g<(\xi-\zeta)$}}                                & \multicolumn{2}{c}{{\rm $dg\nearrow,~dz\nearrow$}}  & \multicolumn{3}{c}{{\rm $dg\nearrow,~dz\searrow$}}  & \multicolumn{2}{c}{{\rm $dz_{\rm cr}\nearrow$}}  \\
\multicolumn{3}{c}{{\rm $(\xi-\zeta)<g$}}                        & \multicolumn{2}{c}{{\rm $dg\searrow,~dz\nearrow$}}  & \multicolumn{3}{c}{{\rm $dg\searrow,~dz\searrow$}}  & \multicolumn{2}{c}{{\rm $dz_{\rm cr}\searrow$}}  \\
\enddata
\tablenotetext{a}{$M_{\rm i}$ -- the initial stellar system's mass. The yields are in solar units ($Z_\mathSun=0.0152$, \citet{gir02}), so that the value $z' = \frac{Z}{p_{\rm eff}} = \frac{Z_\mathSun\cdot10^{\rm [\alpha/H]}}{p_{\rm eff}}\cong\frac{10^{\rm [\alpha/H]}}{p_{\rm eff}}$ used in GCE derivations is unitless.}
\tablenotetext{b}{For ${\rm [\alpha/H]}<{\rm [\alpha/H]}_{\rm s}$ ($\xi-\zeta>0$) we assume the first column of the first row by considering only solutions where $g<\xi$.  After RPS has occurred, we set $\xi=0.01$. This prevents a situation where $dz/dt < 0$ due to intense metal-poor inflow. As a result of our assumptions, for ${\rm [\alpha/H]}>{\rm [\alpha/H]_{\rm s}}$ ($\xi-\zeta<0$) the first column of the second row is used (Section~\ref{sec:A2}).}
\end{deluxetable*}

%% file: results_updated.tab
\begin{deluxetable*}{@{\extracolsep{0pt}}cllllllll}
\label{tab:results}
\tablecolumns{9}
\tablecaption{Parameters of the GCE Models\tablenotemark{a}}
\tablehead{
 & \multicolumn{1}{c}{Leaky Box} & \multicolumn{2}{c}{Pre-enriched} & \multicolumn{3}{c}{RPS} & \multicolumn{2}{c}{Accretion} \\ \cmidrule(lr){2-2} \cmidrule(lr){3-4} \cmidrule(lr){5-7} \cmidrule(lr){8-9}
\multicolumn{1}{c}{Target}  &
\multicolumn{1}{c}{$\peff$}     &
\multicolumn{1}{c}{$\peff$}    &
\multicolumn{1}{c}{${\rm [Fe/H]}_{0}$}     &
\multicolumn{1}{c}{$\peff$}  & 
\multicolumn{1}{c}{${\rm Z}_{\rm s}$ ($10^{-4}$)}    &
\multicolumn{1}{c}{$\zeta$} &
\multicolumn{1}{c}{$\peff$}    &
\multicolumn{1}{c}{$M$}
}
\startdata
Priors & $0 - 100$ & $1 - 25$ & $-5 - -0.5$ & $0 - 5 (1)$ & $1 - 3$ & $0 - 50$ & $0 - 50$    & $1 - 10$ \\
Edges  & $0 - 100$ & $1 - 25$ & $-6 - 0$    & $0 - 5 (1)$ & $1 - 30$ & $0 - 50$ & $0 - 100$   & $1 - 10$ \\
\hline
Draco     & $1.2^{+0.1}_{-0.1}$ & $1.1^{+0.1}_{-0.04}$ & $-3.23^{+0.12}_{-0.15}$ & $0.27^{+0.24}_{-0.14}$ & $3^{+0}_{-1}$ & $>37.2^{+9.2}_{-13.9}$ & $1.2^{+0.1}_{-0.1}$ & $3.29^{+1.13}_{-1.06}$ \\
\hline
Sculptor  & $3.0^{+0.2}_{-0.2}$ & $2.0^{+0.2}_{-0.2}$ & $-2.44^{+0.08}_{-0.09}$ & $0.96^{+0.53}_{-0.42}$ & $6^{+1}_{-1}$ & $>38.5^{+8.3}_{-12.8}$ & $2.9^{+0.1}_{-0.1}$ & $7.98^{+1.32}_{-1.42}$ \\
\hline
Fornax    & $12.0^{+0.6}_{-0.6}$ & $7.9^{+0.5}_{-0.5}$ & $-1.85^{+0.06}_{-0.06}$ & $3.21^{+1.14}_{-1.17}$ & $24^{+3}_{-3}$ & $>36.2^{+9.5}_{-11.7}$ & $11.0^{+0.4}_{-0.4}$ & $7.15^{+1.42}_{-1.24}$ \\
\hline
Leo II    & $2.3^{+0.2}_{-0.2}$ & $2.0^{+0.2}_{-0.2}$ & $-3.18^{+0.24}_{-0.41}$ & $0.49^{+0.35}_{-0.23}$ & $5^{+1}_{-1}$ & $>38.4^{+8.3}_{-13.0}$ & $2.3^{+0.2}_{-0.1}$ & $4.71^{+1.70}_{-1.26}$ \\
\hline
Leo I     & $4.4^{+0.2}_{-0.2}$ & $2.2^{+0.2}_{-0.2}$ & $-2.01^{+0.04}_{-0.05}$ & $0.98^{+0.37}_{-0.29}$ & $10^{+1}_{-1}$ & $>43.2^{+5.0}_{-8.8}$ & $4.3^{+0.1}_{-0.1}$ & $>9.40^{+0.45}_{-0.88}$ \\
\hline
And XVIII & $7.5^{+1.6}_{-1.2}$ & $5.0^{+1.5}_{-1.2}$ & $-1.99^{+0.20}_{-0.32}$ & $2.69^{+1.55}_{-1.68}$ & $13^{+7}_{-6}$ & $28.4^{+13.8}_{-13.1}$ & $6.8^{+1.0}_{-0.8}$ & $>7.05^{+2.03}_{-2.66}$ \\
\hline
\hline
IRPS & Priors & Edges & Draco & Sculptor & Fornax & Leo II & Leo I & And XVIII \\ 
\hline
\multicolumn{1}{c}{$\peff$}                & $0 - 5$       &   $0 - 100$    & $23.6^{+14.8}_{-8.4}$   & $10.4^{+8.4}_{-4.4}$   & $44.4^{+24.0}_{-16.9}$  & $26.9^{+12.4}_{-8.7}$   & $9.6^{+4.1}_{-2.5}$ & $49.2^{+34.0}_{-29.2}$ \\
\hline
\multicolumn{1}{c}{$\zeta/\xi$}            & $0.1-3~(0.5)$ & $0.1-3~(0.5)$  & $>2.47^{+0.38}_{-0.68}$  & $1.08^{+0.75}_{-0.48}$   & $0.83^{+0.38}_{-0.30}$  & $>2.59^{+0.30}_{-0.51}$ & $>2.42^{+0.42}_{-0.65}$  & $0.72^{+0.73}_{-0.40}$  \\
\hline
\multicolumn{1}{c}{${\rm [Fe/H]}_{\rm s}$} & $-3 - -0.85$  &  $-4.5 - -0.7$ & $-1.94^{+0.05}_{-0.12}$ & $-1.52^{+0.04}_{-0.04}$ & $-0.98^{+0.04}_{-0.04}$ & $-1.60^{+0.03}_{-0.04}$ & $-1.29^{+0.01}_{-0.01}$ & $-1.22^{+0.11}_{-0.21}$\\
\hline
\multicolumn{1}{c}{$t_{\rm s}$, Gy ago}    & -- & -- & 11.0 & 11.6 & 7.2 & 6.8 & 1.8 & --\\
\hline
\hline
\multicolumn{3}{c}{Fitting for [$\alpha$/H]} &  \multicolumn{6}{c}{AICc}  \\ 
\hline
\multicolumn{3}{c}{Leaky Box}           & 4624.44 & 5506.73 & 10012.51 & 3905.14 & 11538.33 & 566.11 \\
\multicolumn{3}{c}{Pre-enriched}        & 4604.02 & 5487.50 &  9893.02 & 3897.72 & 11334.23 & 557.13 \\
\multicolumn{3}{c}{Accretion}           & 4597.05 & 5395.54 &  9861.79 & 3858.71 & 11247.81 & 557.03 \\
\multicolumn{3}{c}{RPS}                 & 4591.41 & 5437.38 &  9908.21 & 3847.81 & 11304.71 & 566.91 \\
\multicolumn{3}{c}{IRPS}                & 4588.21 & 5403.12 &  9862.08 & 3839.39 & 11192.62 & 562.35 \\
\multicolumn{3}{c}{Old/new best model}  & RPS/IRPS & Accr/Accr & Accr/Accr & RPS/IRPS & Accr/IRPS & Accr/Accr\\
\enddata
\tablenotetext{a}{$p_{\rm eff}$ is in units of $0.01 Z_{\mathSun}$ ($1Z_{\mathSun}$ for the RPS model). ${\rm Z}_{\rm s} = {\rm Z}_\mathSun 10^{\rm [\alpha/H]}=0.0152\cdot 10^{\rm [\alpha/H]}$ with $Z_\mathSun$ -- the solar metal fraction from~\citet{gir02}. The values in parentheses in the columns titled ``Priors'' and ``Edges'' were used for Sculptor. The error bars illustrate the 68$\%$ confidence interval, while the $>$, $<$ symbols indicate limits.}
\end{deluxetable*}

%% file: targets.tab
\begin{deluxetable*}{cccccccccc}
\label{tab:targets}
\tablecolumns{10}
\tablecaption{Parameters of the Target Galaxies.\tablenotemark{a}}
\tablehead{  & R.A. (J2000) & Decl. (J2000) & D$_{\rm GC}$ & {\rm $L$} & {\rm $r_{1/2}$} & $v_{\mathSun}$  & {\rm $\mu_{\alpha^*}$}  & {\rm $\mu_{\delta}$} & $N_{\rm m}$ \\
Target & & & (kpc) & ($10^5 L_\mathSun$) & (pc) & (km s$^{-1}$) & ($\mu$as yr$^{-1}$) & ($\mu$as yr$^{-1}$) & }
\startdata
Draco     & $17^{\rm h}20^{\rm m}12^{\rm s}.4$ &  $+57^\circ 54' 55''$ & $92\pm29$   & $2.7\pm0.4$ & $220\pm11$        & $-291.0^{+0.1}_{-0.1}$  & $+44^{+5}_{-6}(-94)$     & $-188^{+6}_{-6}(-397)$   & 300 \\
Sculptor  & $01^{\rm h}00^{\rm m}09^{\rm s}.4$ & $-33^\circ 42' 33''$  & $85\pm23$   & $22\pm10$   & $282\pm41$        & $111.4^{+0.1}_{-0.1}$   & $+100^{+2}_{-2}(+96)$    & $-158^{+2}_{-2}(-149)$   & 370 \\
Fornax    & $02^{\rm h}39^{\rm m}59^{\rm s}.3$ & $-34^\circ 26' 57''$  & $141\pm12$  & $180\pm50$  & $714\pm40$        & $55.3^{+0.1}_{-0.1}$    & $+381^{+1}_{-1}(+380)$   & $-359^{+2}_{-2}(-308)$   & 665 \\
Leo II    & $11^{\rm h}13^{\rm m}28^{\rm s}.8$ & $+22^\circ 09' 06''$  & $221\pm50$  & $6.6\pm1.9$ & $177\pm13$        & $78.0^{+0.1}_{-0.1}$    & $-109^{+28}_{-28}(-276)$ & $-150^{+26}_{-26}(-255)$  & 259 \\
Leo I     & $10^{\rm h}08^{\rm m}28^{\rm s}.1$ & $+12^\circ 18' 23''$  & $257\pm76$  & $56\pm16$   & $295\pm49$        & $282.5^{+0.1}_{-0.1}$   & $-50^{+14}_{-14}(-66)$   & $-120^{+10}_{-10}(-121)$ & 777 \\
And XVIII & $00^{\rm h}02^{\rm m}14^{\rm s}.5$ & $+45^\circ 05' 20''$  & $579\pm87$  & $6.2\pm0.4$ & $356^{+35}_{-47}$ & $-337.2^{+1.5}_{-1.4}$  & --                       & --                       & 38 \\
\enddata
\tablenotetext{a}{The parameters are from \citet{kir11a}, \citet{mcc12}, \citet{kva24}, and references therein. $D_{\rm GC}$ is the distance to the host; {\rm $r_{1/2}$} is a two-dimensional projected half-light radius. The proper motions are from \citet[][values for member stars $N_{\rm m}$ are in parentheses]{pac22}.}
\end{deluxetable*}